\documentclass[
    ,draft            
  ]
  {aipproc}

\layoutstyle{6x9}

\newcommand{\Mdot}{\dot{M}}
\newcommand{\Mwd}{M_{\rm wd}}
\newcommand{\Rwd}{R_{\rm wd}}

\newcommand{\lax}{{\lower0.75ex\hbox{ $<$ }\atop\raise0.5ex\hbox{ $\sim$ }}}
\newcommand{\gax}{{\lower0.75ex\hbox{ $>$ }\atop\raise0.5ex\hbox{ $\sim$ }}}
\newcommand{\ion}[2]{#1~{\sc\uppercase\expandafter{\romannumeral #2}}}

\begin{document}

\title{Fe L-Shell Density Diagnostics in Theory and Practice}




\author{Christopher W.~Mauche}{
  address={Lawrence Livermore National Laboratory,
   L-473, 7000 East Avenue, Livermore, CA 94550},
}

\author{Duane A.~Liedahl}{
  address={Lawrence Livermore National Laboratory,
   L-473, 7000 East Avenue, Livermore, CA 94550},
}

\author{Kevin B.~Fournier}{
  address={Lawrence Livermore National Laboratory,
   L-473, 7000 East Avenue, Livermore, CA 94550},
}

\begin{abstract}

We provide a discussion of the density and photoexcitation sensitivity
of the X-ray spectra of Fe L-shell ions (\ion{Fe}{17}--\ion{Fe}{24})
calculated with the Livermore X-ray Spectral Synthesizer, a suite of
IDL codes that calculates spectral models of highly charged ions
based primarily on HULLAC atomic data. These models are applicable
to collisionally ionized laboratory or cosmic plasmas with electron
temperatures $T_{\rm e}\approx 2$--45 MK (0.2--4 keV) and electron
densities $n_{\rm e}\gax 10^{11}~\rm cm^{-3}$. Potentially useful
density diagnostics are identified for \ion{Fe}{17} and
\ion{Fe}{19}--\ion{Fe}{23}, with the most straightforward being the
\ion{Fe}{17} $I$(17.10 \AA )/$I$(17.05 \AA ) line ratio and the
\ion{Fe}{22} $I$(11.92 \AA )/$I$(11.77 \AA ) line ratio.
Applying these models to the {\it Chandra\/} X-ray Observatory High
Energy Transmission Grating spectrum of the intermediate polar EX~Hya,
we find that the strength of all the Fe L-shell lines are consistent
with electron densities $n_{\rm e}\gax 1\times 10^{14}~\rm cm^{-3}$.
Specifically, from the observed \ion{Fe}{17} $I$(17.10 \AA )/$I$(17.05
\AA ) line ratio, we infer an electron density $n_{\rm e}\gax 2\times
10^{14}~\rm cm^{-3}$ at the $3\, \sigma $ level, while from the observed
\ion{Fe}{22} $I$(11.92 \AA )/$I$(11.77 \AA ) line ratio, we infer
$n_{\rm e}=1.0^{+2.0}_{-0.5} \times 10^{14}~\rm cm^{-3}$ at the $1\,
\sigma $ level and $n_{\rm e} \gax 2\times 10^{13}~\rm  cm^{-3}$ at
the $3\, \sigma $ level.

\end{abstract}

\maketitle


\section{Introduction}

The standard density diagnostic of high-temperature plasmas is the
intensity ratio $R\equiv f/i$ of the forbidden to intercombination 
lines of He-like ions, which falls from its low-density value $R_0$
to zero at a critical density (where $R/R_0=0.5$) that increases from
$n_{\rm e}\approx 3\times 10^{8}~\rm cm^{-3}$ for C to $n_{\rm e}
\approx 3 \times 10^{17}~\rm cm^{-3}$ for Fe \citep{gab69, blu72, por01,
por05} (see left panel of Fig~1). Unfortunately, this diagnostic is
compromised in ultraviolet-bright sources like O stars and cataclysmic
variables, wherein photoexcitation competes with collisional excitation
to depopulate the upper level of the forbidden line, causing the $R$
line ratio to appear to be in the high-density limit regardless of the
density; for a plasma illuminated by a 30 kK blackbody, this is true for
all elements through Mg \citep{mau02} (see right panel of Fig~1). In an
attempt to circumvent this problem, we have undertaken an investigation
of potential density diagnostics of lines of Fe L-shell ions
(\ion{Fe}{17}--\ion{Fe}{24}). These diagnostics are applicable to
collisionally ionized plasmas with electron temperatures $T_{\rm e}
\approx 2$--45 MK (0.2--4 keV) and electron densities $n_{\rm e}\gax 
10^{11}~\rm cm^{-3}$. Given the high densities, these
diagnostics are particularly suitable to magnetic cataclysmic variables
(polars and intermediate polars \citep{war95}), in which the density of
the X-ray--emitting plasma is increased by (1) the magnetic channeling
of the mass lost by the secondary onto small spots near the white dwarf
magnetic poles; (2) the factor-of-four density jump across the standoff
shock, which heats the plasma to a temperature $T_{\rm s} =3G\Mwd\mu
m_{\rm H}/8k\Rwd \approx 200$~MK; and (3) the settling nature of the
post-shock flow, wherein the pressure is roughly constant and the
density scales inversely with the decreasing temperature \citep{fra92}.
For a mass-accretion rate $\Mdot = 10^{15}~\rm g~s^{-1}$ (hence luminosity
$L=G\Mwd\Mdot /\Rwd\approx 6\times 10^{31}~\rm erg~s^{-1}$), a free-fall
velocity $v_{\rm ff} = (2G\Mwd/\Rwd)^{1/2} \approx 3600~\rm km~s^{-1}$, 
and a relative spot size $f=0.1$, the density of the accretion flow
immediately behind the shock $n=\Mdot/ 4\pi f\Rwd ^2\mu m_{\rm H}
(v_{\rm ff}/4)\approx 10^{13}~\rm cm^{-3}$.

In two previous publications, we explored the density and
photoexcitation sensitivity of the
\ion{Fe}{17} $I$(17.10 \AA )/$I$(17.05 \AA ) line ratio \citep{mau01}
and the
\ion{Fe}{22} $I$(11.92 \AA )/$I$(11.77 \AA ) line ratio \citep{mau03} 
and applied them to the {\it Chandra\/} X-ray Observatory High-Energy
Transmission Grating Spectrometer (HETGS) Medium Energy Grating (MEG)
spectrum of the intermediate polar EX Hya \citep{gre97, hur97, fuj97,
eis02, hoo04}. We summarize the results of those investigations below,
but first provide a general look at the {\it Chandra\/} HETG spectrum
of EX Hya and the density sensitivity of the X-ray spectra of all Fe
L-shell ions.

\begin{figure}
\includegraphics{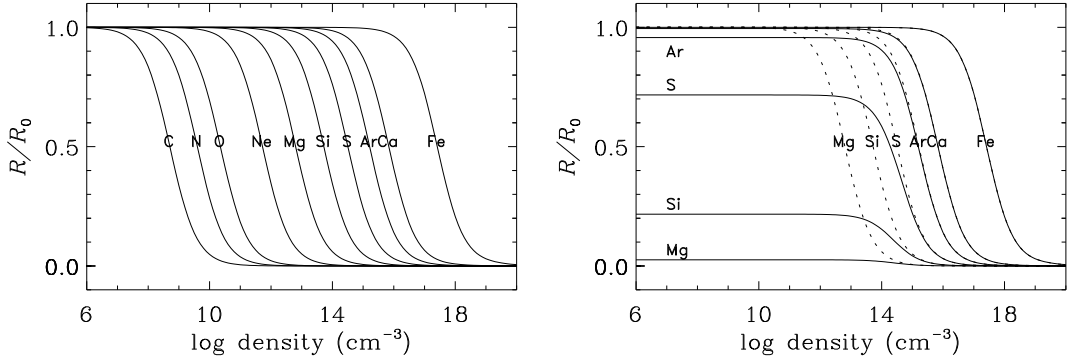}
\caption{{\it Left panel\/}: He-like $R\equiv z/(x+y)=f/i$ line ratios
of the abundant elements as a function of electron density. For C, N,
O, \ldots , Fe, the plasma temperatures $T_{\rm e}= 0.46$, 0.57, 0.86,
\ldots, 36.1 MK, the temperature of the peak ionization fraction for
each ion \citep{maz98}.
{\it Right panel\/}: Similar to the left panel ({\it dotted curves\/}), 
but accounts for photoexcitation by a $T_{\rm bb}=30$ kK blackbody
({\it full curves\/}). In both panels the line ratios are scaled to the
low-density limit $R_0$ for ease of comparison.}
\label{fig1}
\end{figure}

\section{Chandra Spectrum of EX Hya}

We begin by comparing the {\it Chandra\/} MEG spectrum of EX Hya
with that of the coronally active RS CVn binary HR 1099 \citep{ayr01,
nes02, ost04}. As one of the targets of the {\it Chandra\/} Emission
Line Project (\url{http://asc.harvard.edu/elp/ELP.html}) \citep{bri00},
HR 1099 was observed extensively by {\it Chandra\/}, and its X-ray
spectrum is one of the benchmarks for models of collisionally
ionized plasmas. Figure~2 shows the MEG count spectra of EX Hya
({\it blue histogram\/}) and HR 1099 ({\it red histogram\/}) in the
10.5--17.5~\AA \ bandpass, which contains the strongest lines of Fe
L-shell ions. The spectrum of EX Hya differs from that of HR 1099 in
(1) the absence of the     \ion{Fe}{17} 17.10~\AA \ line,
(2) the strength of the    \ion{Fe}{17} 17.05~\AA \ line,
(3) the absence of the     \ion{Ne}{9}  13.70~\AA \ forbidden line,
(4) the weakness of the    \ion{Fe}{20} 12.83~\AA \ line,
(5) the absence of the     \ion{Fe}{21} 12.26~\AA \ line, and
(6) the strength of the    \ion{Fe}{22} 11.92~\AA \ line.
As shown by Fig~1, the absence of the \ion{Ne}{9} forbidden line
is indicative of electron densities $n_{\rm e}\gax 1\times 10^{13}~\rm
cm^{-3}$ or of photoexcitation by a $T_{\rm bb}\gax 30$ kK blackbody.
The later possibility cannot be excluded in EX Hya, whose far
ultraviolet spectrum has been modeled as a $T_{\rm bb}\approx 25$~kK
white dwarf \citep{gre97} and as a $T_{\rm bb}\approx 20$~kK white
dwarf with a $T_{\rm bb}\approx 37$~kK hot spot \citep{mau99}. As we will
see below, the  \ion{Fe}{17} $I$(17.10 \AA )/$I$(17.05 \AA ) line ratio
and the \ion{Fe}{22} $I$(11.92 \AA )/$I$(11.77 \AA ) line ratio are
less sensitive to photoexcitation, and indeed the strength of all the
Fe L-shell lines are consistent with electron densities $n_{\rm e}\gax
1\times 10^{14}~\rm cm^{-3}$.

\begin{figure}
\includegraphics{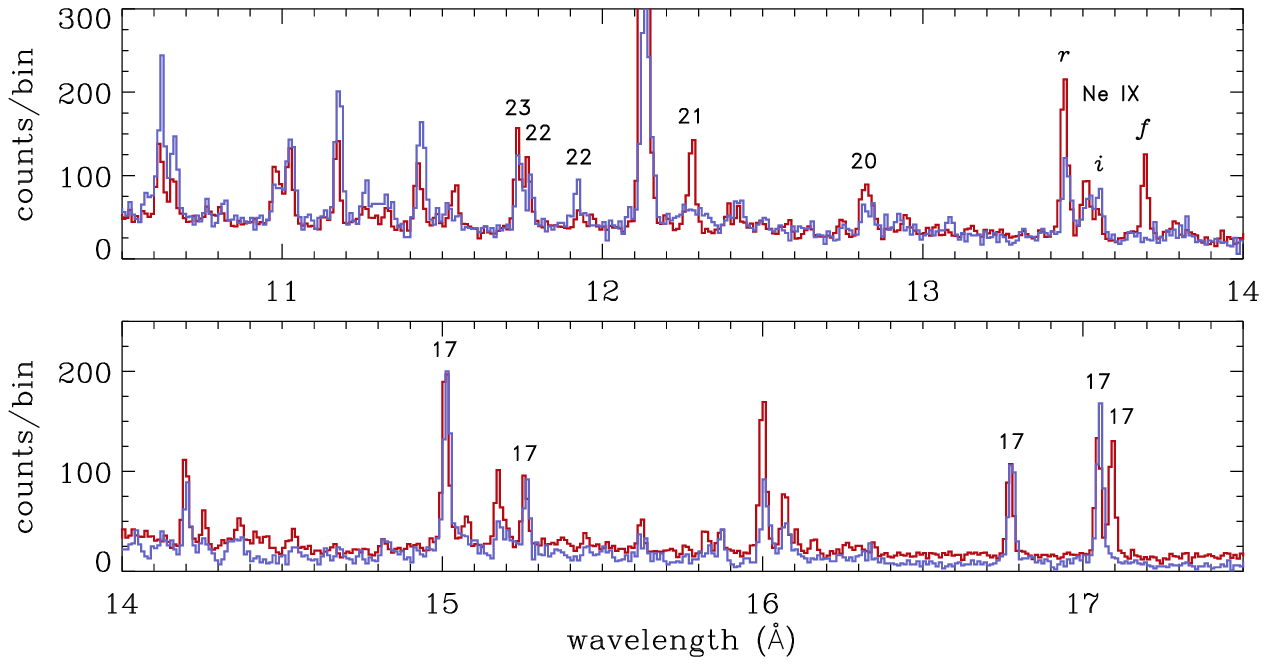}
\caption{{\it Chandra\/} MEG count spectra (combining $\pm 1$st orders
and binned to 0.005~\AA ) of EX Hya ({\it blue histogram\/}) and HR1099
({\it red histogram\/}). The HR 1099 data is scaled by a factor of
0.25 (0.34) in the upper (lower) panel, and the \ion{Ne}{9} and Fe
L-shell lines discussed in the text are labeled.}
\label{fig2}
\end{figure}

\section{LXSS Fe L-Shell Model Spectra}

We constructed model spectra of Fe L-shell ions using the Livermore
X-ray Spectral Synthesizer (LXSS), a suite of IDL codes that
calculates spectral models of highly charged ions based on Hebrew
University/Lawrence Livermore Atomic Code (HULLAC) atomic data. HULLAC
calculates atomic wavefunctions, level energies, radiative transition
rates, and oscillator strengths according to the fully relativistic,
multiconfiguration, parametric potential method \citep{kla71, kla77}.
Electron impact excitation rate coefficients are computed
quasi-relativistically in the distorted wave approximation \citep{bar88}
assuming a Maxwellian velocity distribution. Table~1 lists the number of
energy levels, radiative transition rates for E1, E2, M1, and M2 decays,
and electron impact excitation rate coefficients for each of our Fe
L-shell models. Using these data, LXSS calculates the level populations
for a given temperature and density assuming collisional-radiative
equilibrium. The line intensities are then simply the product of the
level populations and the radiative transition rates. In the sections
below on \ion{Fe}{17} and \ion{Fe}{22}, we account for photoexcitation by
including in the LXSS population kinetics calculation the photoexcitation
rates $(\pi e^2/m_ec) f_{ij} F_\nu (T)$, where $F_\nu (T)$ is the
continuum spectral energy distribution and $f_{ij}$ are the oscillator
strengths of the various transitions. For simplicity, we assume that
$F_\nu (T) = (4\pi /h\nu ) B_\nu (T_{\rm bb})$ (i.e., the radiation field
is that of a blackbody of temperature $T_{\rm bb}$) and the dilution
factor of the radiation field is equal to $1\over 2$ (i.e., the X-ray
emitting plasma is in close proximity to the source of the
photoexcitation continuum).

\begin{table}[!t]
\begin{tabular}{lrrr}
\hline
\tablehead{1}{l}{b}{Ion}    &
\tablehead{1}{c}{b}{Levels} &
\tablehead{1}{c}{b}{Collisional\\Rates} &
\tablehead{1}{c}{b}{Radiative\\Rates} \\
\hline
\ion{Fe}{17} & 281 &  33{,}887 &  49{,}882 \\
\ion{Fe}{18} & 456 &  93{,}583 & 141{,}229 \\
\ion{Fe}{19} & 605 & 164{,}496 & 240{,}948 \\
\ion{Fe}{20} & 609 & 165{,}350 & 257{,}765 \\
\ion{Fe}{21} & 591 & 153{,}953 & 227{,}743 \\
\ion{Fe}{22} & 228 &  24{,}084 &  37{,}300 \\
\ion{Fe}{23} & 116 &   6{,}478 &   8{,}798 \\
\ion{Fe}{24} &  76 &   1{,}704 &   4{,}100 \\
\hline
\end{tabular}
\caption{LXSS Fe L-shell Models}
\label{tab1}
\end{table}

The resulting Fe L-shell ion spectra are shown in Figs 3--6. In each
case, the atomic model was calculated at the temperature of the peak
ionization fraction for each ion \citep{maz98} for electron densities
$n_{\rm e}=10^{10}$--$10^{18}~\rm cm^{-3}$, and the spectra were binned to
0.02 \AA , which is approximately equal to the FWHM of the MEG spectra.
The upper and lower panels of each figure show the model spectra  at
the extremes of the density grid: $n_{\rm e} = 10^{10}~\rm cm^{-3}$ in
red and $n_{\rm e}=10^{18}~\rm cm^{-3}$ in blue. Note that each panel
encompasses a different 4~\AA \ bandpass, allowing them to be compared
directly to Fig~2. In each case, the spectrum is scaled to the
brightest line in the $n_{\rm e}= 10^{10}~\rm cm^{-3}$ model and by a
factor of $(10^{10}~{\rm cm^{-3}})/n_{\rm e}$. The middle panels of each
figure show as a function of electron density the relative line strengths
for lines brighter than 0.25 photons/bin. Each of these models is
discussed in turn.

\begin{figure}
\includegraphics{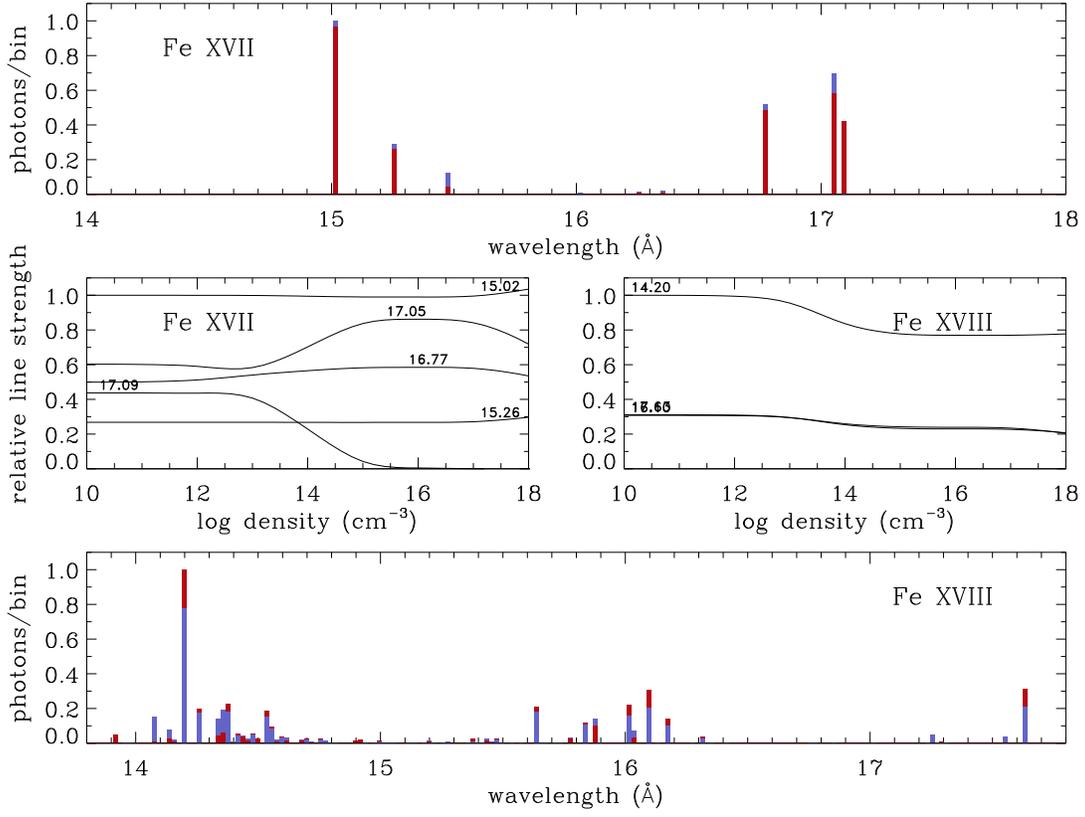}
\caption{LXSS models of \ion{Fe}{17} at $T_{\rm e}=4.1$ MK and
\ion{Fe}{18} at $T_{\rm e}=6.6$ MK. Upper and lower panels show the
model spectra binned to 0.02~\AA \ for $n_{\rm e}=10^{10}~\rm cm^{-3}$
({\it red bars\/}) and $n_{\rm e}=10^{18}~\rm cm^{-3}$ ({\it blue
bars\/}); middle panels show as a function of electron density the
relative line strengths for lines brighter than 0.25 photons/bin.}
\label{fig3}
\end{figure}

\begin{figure}
\includegraphics{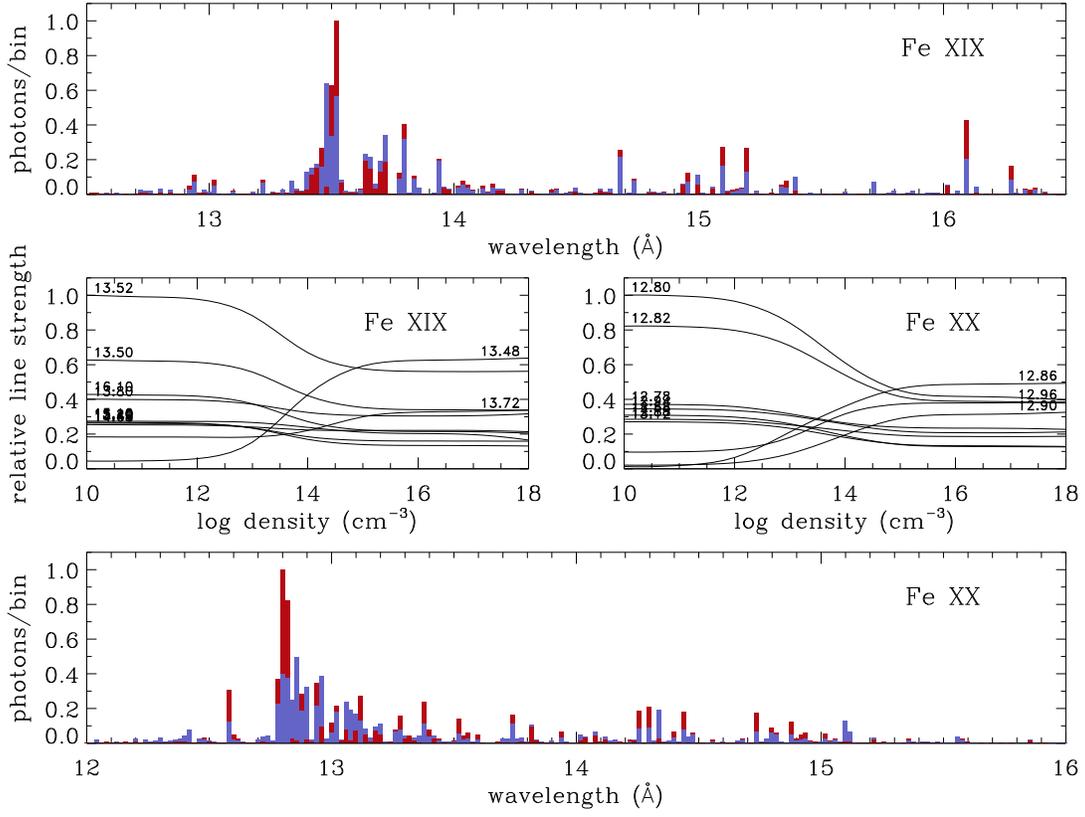}
\caption{Similar to Fig~2, for \ion{Fe}{19} at $T_{\rm e}=7.9$ MK and
\ion{Fe}{20} at $T_{\rm e}=9.5$ MK.}
\label{fig4}
\end{figure}

\begin{figure}
\includegraphics{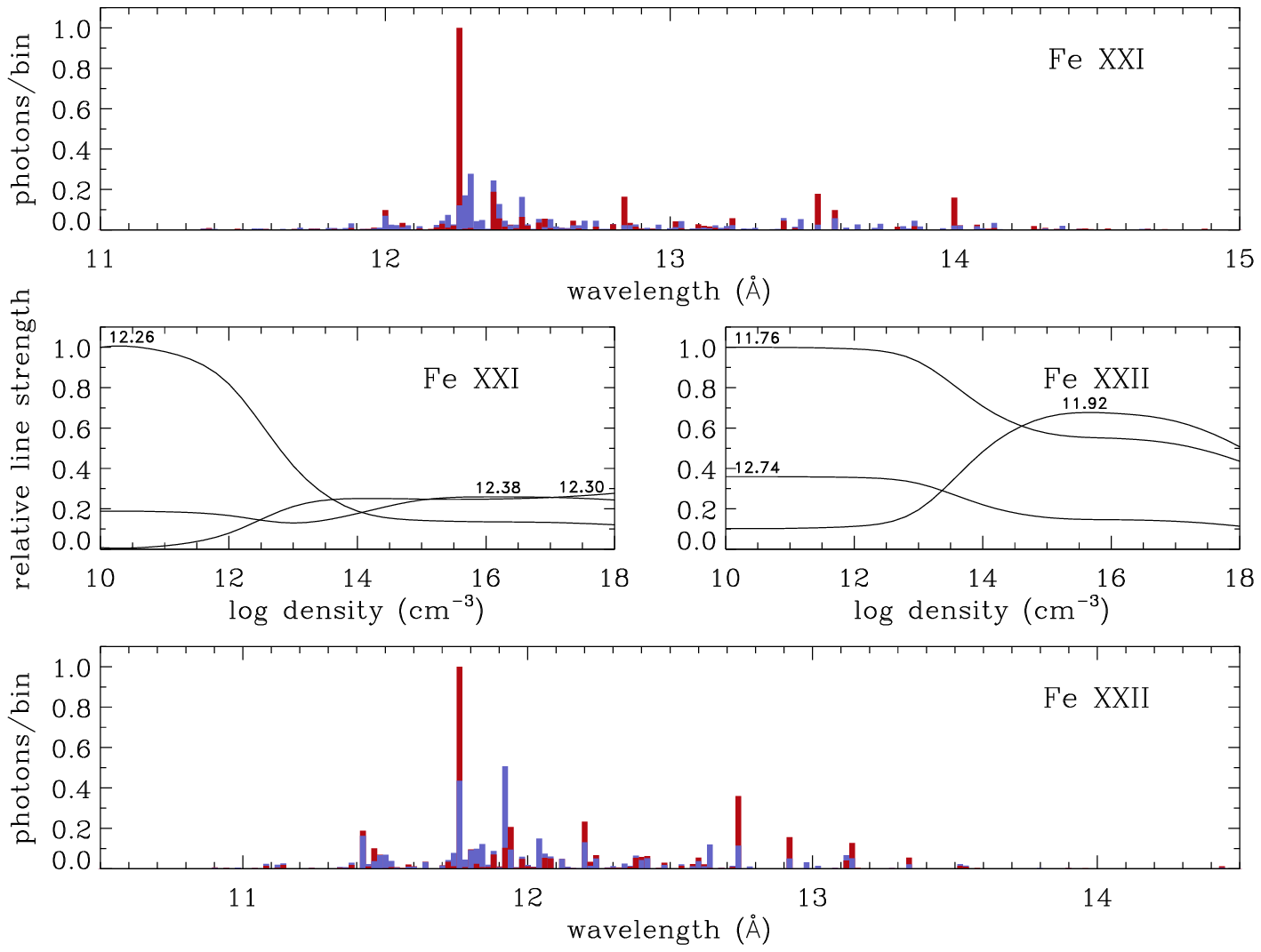}
\caption{Similar to Fig~2, for \ion{Fe}{21} at $T_{\rm e}=10.5$ MK and
\ion{Fe}{22} at $T_{\rm e}=12.3$ MK.}
\label{fig5}
\end{figure}

\begin{figure}
\includegraphics{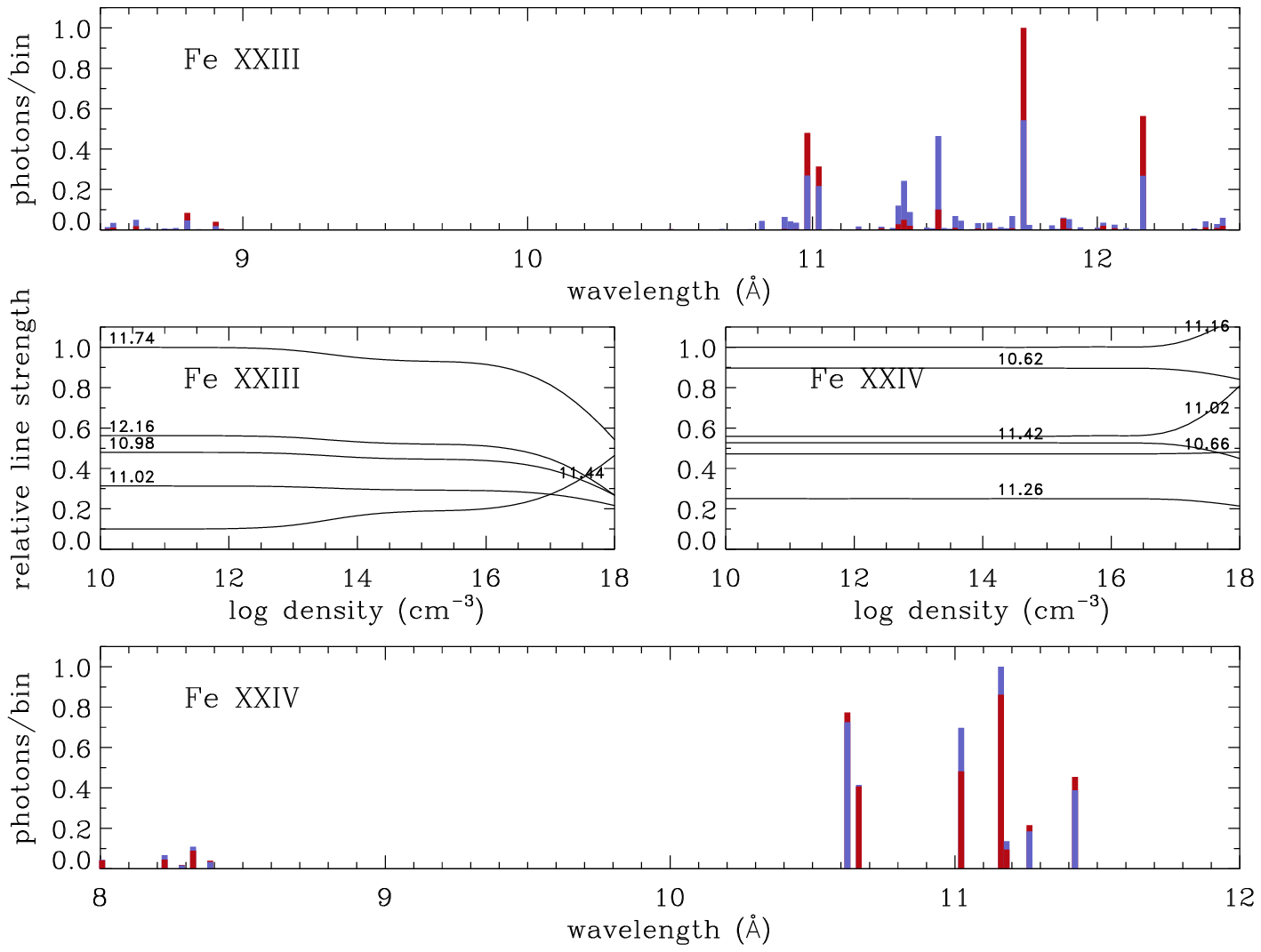}
\caption{Similar to Fig~2, for \ion{Fe}{23} at $T_{\rm e}=14.1$ MK and
\ion{Fe}{24} at $T_{\rm e}=17.8$ MK.}
\label{fig6}
\end{figure}

\smallskip \noindent {\bf Fe~XVII}:
The X-ray spectrum of Ne-like \ion{Fe}{17} is dominated at low densities
by the three $3d$--$2p$ lines near 15~\AA \ and the three $3s$--$2p$
lines near 17~\AA . As the density increases, the 17.10~\AA \ line
disappears and the 17.05~\AA \ line brightens. As shown below,
the $I$(17.10 \AA )/$I$(17.05 \AA ) line ratio observed in EX Hya is
consistent with $n_{\rm e} \gax 2 \times 10^{14}~\rm cm^{-3}$ at
$T_{\rm e}\approx 4.1$ MK.

\smallskip \noindent {\bf Fe~XVIII}:
The X-ray spectrum of F-like \ion{Fe}{18} is dominated at low densities
by the $3d$--$2p$ line at 14.20~\AA . As the density increases, the
three brightest lines of this ion all dim, providing no useful density
diagnostic.

\smallskip \noindent {\bf Fe~XIX}:
The X-ray spectrum of O-like \ion{Fe}{19} is dominated at low densities
by a tight clump of $3d$--$2p$ lines near 13.5~\AA . As the density
increases, the 13.52~\AA \ and 13.50~\AA \ lines dim and the 13.48~\AA \
line brightens, providing a potential density diagnostic, although these
lines are blended with the (typically brighter) He $\alpha $ lines of
\ion{Ne}{9}.

\smallskip \noindent {\bf Fe~XX}:
The X-ray spectrum of N-like \ion{Fe}{20} is dominated at low densities
by a broad clump of $3d$--$2p$ lines near 12.8~\AA . As the density
increases, the 12.80~\AA \ and 12.82~\AA \ lines dim and the 12.86~\AA ,
12.96~\AA , and 12.90~\AA \ lines brighten, providing a potential density
diagnostic. The weakness of the 12.8~\AA \ line in EX Hya is consistent
with $n_{\rm e} \gax 3 \times 10^{14}~\rm cm^{-3}$ at $T_{\rm e}\approx
9.5$ MK.

\smallskip \noindent {\bf Fe~XXI}:
The relatively simple X-ray spectrum of C-like \ion{Fe}{21} is dominated
at low densities by a single $3d$--$2p$ line at 12.26~\AA . As the
density increases, this line dims and the 12.38~\AA \ line brightens,
providing a potential density diagnostic. The absence of the 12.26~\AA \
line in EX Hya is consistent with $n_{\rm e} \gax 3 \times 10^{13}~\rm
cm^{-3}$ at $T_{\rm e}\approx 10.5$ MK.

\smallskip \noindent {\bf Fe~XXII}:
The X-ray spectrum of B-like \ion{Fe}{22} is dominated at low densities
by the $3d_{3/2}$--$2p_{1/2}$ line at 11.77~\AA . As the density
increases, this line dims and the $3d_{5/2}$--$2p_{3/2}$ line at
11.92~\AA \ line brightens, providing a clear density diagnostic. As
shown below, the $I$(11.92 \AA )/$I$(11.77 \AA ) line ratio observed in
EX Hya is consistent $n_{\rm e}\approx 1\times 10^{14}~\rm cm^{-3}$ at
$T_{\rm e}\approx 12.3$ MK.

\smallskip \noindent {\bf Fe~XXIII}:
The X-ray spectrum of Be-like \ion{Fe}{23} consists of a few strong
$3d$--$2p$ lines between 11.0--12.2~\AA . As the densities increases,
the 11.74~\AA \ line dims and the 11.44~\AA \ line brightens, but they
do not become comparable in strength until $n_{\rm e}\approx 10^{18}~\rm
cm^{-3}$.

\smallskip \noindent {\bf Fe~XXIV}:
The X-ray spectrum of Li-like \ion{Fe}{24} consists of a few strong
$3s$--$2p$,  $3d$--$2p$, and $3p$--$2s$ lines between 10.6--11.4~\AA .
There is little density dependence to its X-ray spectrum.

\smallskip \noindent 	
Based on the foregoing, it is clear that the \ion{Fe}{17}
$I$(17.10 \AA )/$I$(17.05 \AA ) line ratio and the \ion{Fe}{22}
$I$(11.92 \AA )/$I$(11.77 \AA ) line ratio provide the most
straightforward density diagnostics of Fe L-shell ions. Each of
these is discussed in more detail below.

\section{Fe~XVII}

Because of the persistence of Ne-like \ion{Fe}{17} over a broad
temperature range ($T_{\rm e}\approx 2$--12 MK) and the large collision
strengths for $2p\to nd$ transitions, the $2p^6$--$2p^53l$ ($l=s$, $d$)
lines of \ion{Fe}{17} at 15--17~\AA \ are prominent in the X-ray spectra
of high-temperature plasmas in tokamaks \citep{kla78, phi97, bei01}, the
Sun \citep{rug85, phi97, sab99}, and both late- and early-type stars
\citep{can00, ayr01, kah01, wal01}. The lowest-lying configurations of
\ion{Fe}{17} are the $2p^6$ ground state, 
the $2p^53s$ manifold (4  levels),
the $2p^53p$ manifold (10 levels), and
the $2p^53d$ manifold (12 levels) (Fig~7).
Because electron impact excitation from the ground state is strong for 
transitions into the $2p^5nd$ configurations, but weak for transitions
into the $2p^53s$ configuration, the population flux into the 
lowest-lying configuration is dominated by radiative cascades originating
on higher-lying energy levels, rather than direct excitation from the
ground state. The four levels of the $2p^53s$ configuration
have, in increasing energy order, total angular momenta $J=2$, 1, 0, and
1. The $2p^6$--$2p^5 3s\> (J=0)$ transition is strictly forbidden, but
the remaining three levels $J=2$, 1, and 1 decay to ground, producing
lines at 17.10, 17.05, and 16.78~\AA , respectively. The 17.10~\AA \ line
is produced by an M2 transition, but it is nevertheless bright because
the upper level is populated efficiently by radiative cascades, and its
radiative branching ratio to ground is 1.0. Since the radiative decay
rate of this transition is slow compared to the other $2p$--$3l$
lines, collisional depopulation sets in at lower densities. Thus, the
intensity ratio of the 17.10~\AA\ line to any of the other $2p$--$3l$
lines provides a density diagnostic, as was first pointed out by
\citet{kla78}. 

To investigate quantitatively the density and photoexcitation dependence
of the \ion{Fe}{17} $I$(17.10 \AA )/$I$(17.05 \AA ) line ratio, we
calculated LXSS models for this ion for $T_{\rm e}=4.1$ MK, $n_{\rm e}=
10^{10}$--$10^{18}~\rm cm^{-3}$, and $T_{\rm bb}=0$--60 kK (for
additional details, see \citep{mau01}). Figure~8 shows the LXSS model
$I$(17.10 \AA )/$I$(17.05 \AA ) line ratio, the {\it Chandra\/} MEG
spectrum of EX Hya in the neighborhood of the \ion{Fe}{17} $2p$--$3s$
lines, and the value and 1, 2, and $3\sigma $ error envelops of the
measured $I$(17.10 \AA )/$I$(17.05 \AA ) line ratio. The calculation
shows that the blackbody temperature must be $T_{\rm bb} \gax 60$ kK
to drive the $I$(17.10 \AA )/$I$(17.05 \AA ) line ratio into the
high-density limit, while the figure shows that, in EX Hya, $n_{\rm e}
\gax 2\times 10^{14}~\rm cm^{-3}$ or $T_{\rm bb}\gax 50$ kK at the $3\,
\sigma $ level. 

\section{Fe~XXII}

The density sensitivity of the X-ray lines of B-like \ion{Fe}{22} has
been discussed extensively in studies of the X-ray spectra of solar and
laboratory high-temperature plasmas \citep{dos73, mas80, dos81, phi82,
faw87, phi96, war98, che04}. At low densities, the \ion{Fe}{22} electron
population is primarily in the $2s^22p_{1/2}$ ground state, and
collisional excitations are predominantly from the ground state into the
$2s2p^2$ manifold (8 levels) and the $2s^23d_{3/2}$ level, both of which
decay primarily to ground, producing lines in the extreme ultraviolet
(EUV) and at 11.77~\AA , respectively (Fig~7). However, these levels
also decay to the $2s^22p_{3/2}$ first-excited level with approximately
15\% probability, producing lines in the EUV and at 11.92~\AA ; when
{\it that\/} level decays to ground, it produces a line in the
ultraviolet. As the density increases, electron population builds
up in the $2s^22p_{3/2}$ first-excited level because the M1 transition
to ground is slow. At high densities, the $2s^22p_{3/2}$ first-excited
level is fed primarily by radiative decays from the $2s2p^2$ manifold.
Collisional excitations out of the first-excited level are primarily into
the $2s^23d_{5/2}$ level, which decays primarily back to the first-excited
level, producing a line at 11.92~\AA . Consequently, the \ion{Fe}{22}
11.92~\AA \ line is relatively strong in the X-ray spectra of
high-density plasmas.

To investigate quantitatively the density and photoexcitation dependence
of the \ion{Fe}{22} $I$(11.92 \AA )/$I$(11.77 \AA ) line ratio, we
calculated LXSS models for this ion for $T_{\rm e}=12.8$ MK, $n_{\rm e}
=10^{10}$--$10^{18}~\rm cm^{-3}$, and $T_{\rm bb}=0$--100 kK. Motivated
by the detailed study of \citet{fou01} of the density-sensitive Fe
L-shell lines in the EUV, in these models we made two modifications to
the collisional excitation data used in LXSS. First, we replaced, for all
transitions between and among the $2s^22p$ and $2s2p^2$ levels of
\ion{Fe}{22}, the electron impact excitation rate coefficients computed
with HULLAC with those of \citet{zha97} computed with the relativistic
R-matrix method. Second, we added proton excitations for transitions
among the levels of the $2s^22p$ and $2s2p^2$ configuration using the
proton impact excitation rate coefficients of \citet{fos97}.  For
additional details, see \citep{mau03}.

Figure~9 shows the resulting LXSS model $I$(11.92 \AA )/$I$(11.77 \AA )
line ratio, the {\it Chandra\/} MEG spectrum of EX Hya in the neighborhood
of the \ion{Fe}{22} $2p$--$3d$ lines, and the value and 1, 2, and
$3\sigma $ error envelops of the measured $I$(11.92 \AA )/$I$(11.77 \AA )
line ratio. The figure shows that the $I$(11.92 \AA )/$I$(11.77 \AA )
line ratio is relatively insensitive to photoexcitation, and that, in
EX Hya, $n_{\rm e}= 1.0^{+2.0}_{-0.5}\times 10^{14}~\rm cm^{-3}$ at the
$1\, \sigma $ level and $n_{\rm e}\gax 2\times 10^{13}~\rm cm^{-3}$ or
$T_{\rm bb}\gax 100$ kK at the $3\, \sigma $ level.

\begin{figure}
\includegraphics{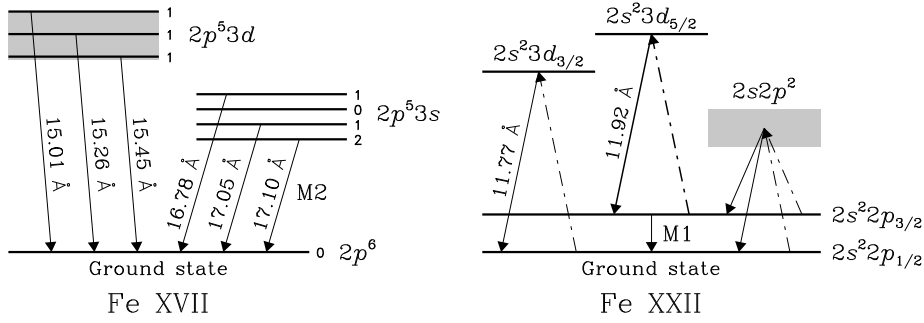}
\caption{%
Simplified Grotrian diagrams for \ion{Fe}{17} and \ion{Fe}{22}.
Collisional channels are shown with dot-dashed lines and radiative
channels with solid lines; X-ray transitions are labeled with the
wavelength.}
\label{fig7}
\end{figure}

\begin{figure}
\includegraphics{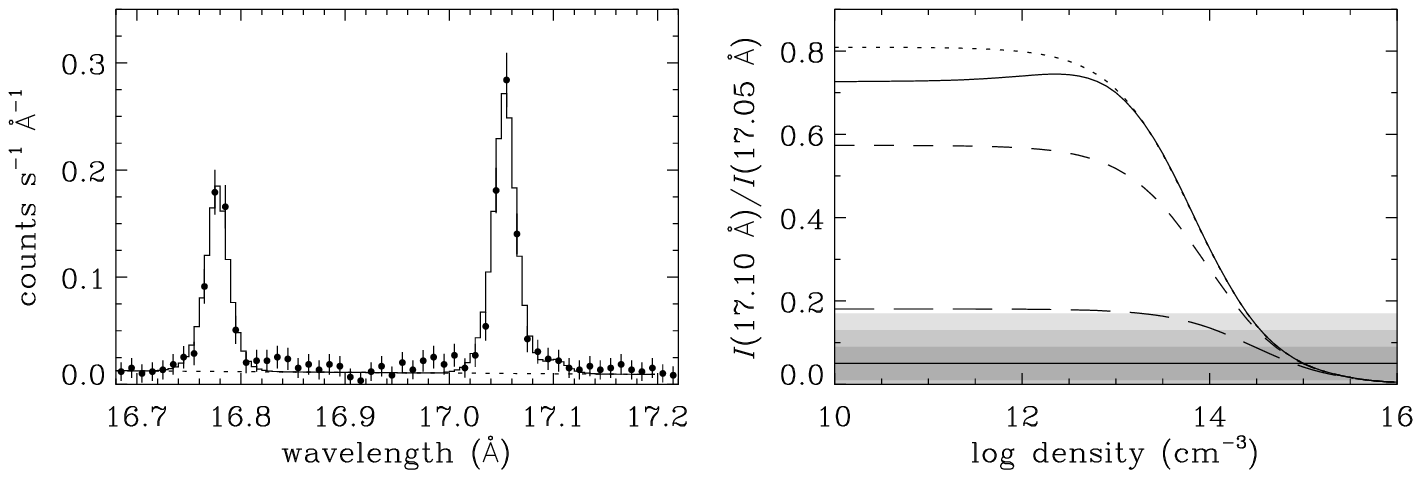}
\caption{%
{\it Left panel\/}: Detail of the {\it Chandra\/} MEG spectrum of EX Hya
in the neighborhood of the \ion{Fe}{17} $2p$--$3s$ lines. Data
(combining $\pm 1$st orders and binned to 0.01~\AA ) are shown by the
filled circles with error bars and the model fit is shown by the
histogram. 
{\it Right panel\/}: LXSS model \ion{Fe}{17} $I$(17.10 \AA )/$I$(17.05
\AA ) line ratio as a function of electron density for $T_{\rm e}=4.1$ MK
and $T_{\rm bb}=10$, 30, 40, and 50 kK ({\it solid, dotted, short dashed,
and long dashed curves, respectively\/}). Horizontal line and the dark,
medium, and light shaded stripes indicate respectively the value and the
68\%, 90\%, and 99\% confidence error envelopes of the line ratio
measured in EX Hya.}
\label{fig8}
\end{figure}

\begin{figure}
\includegraphics{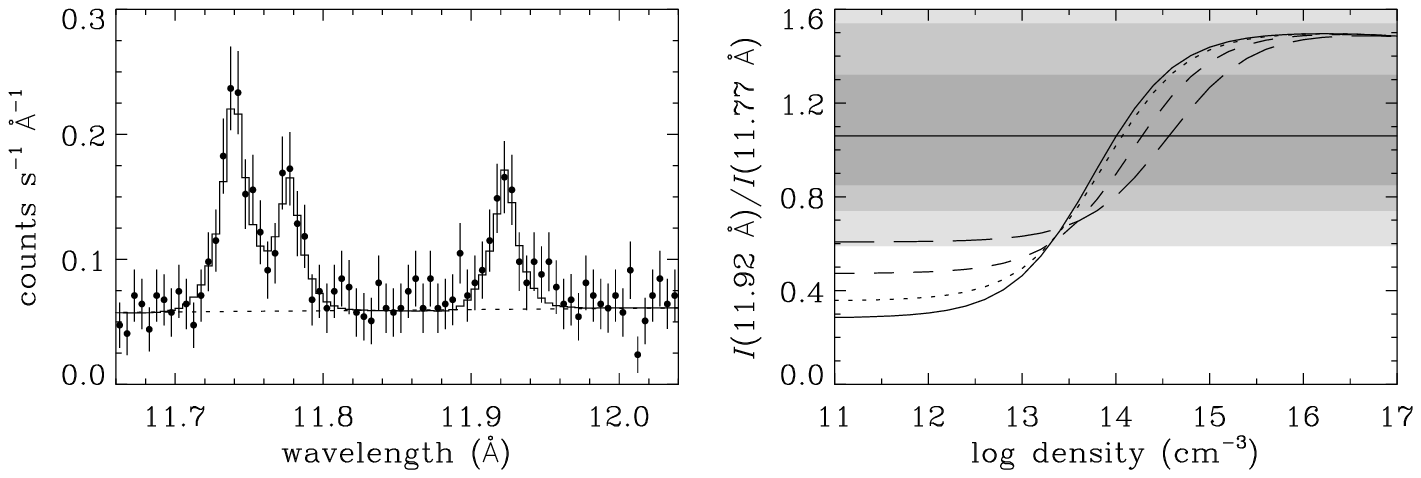}
\caption{%
{\it Left panel\/}: Detail of the {\it Chandra\/} MEG spectrum of EX
Hya in the neighborhood of the \ion{Fe}{22} $2p$--$3d$ lines. Data
(combining $\pm 1$st orders and binned to 0.005~\AA ) are shown by the
filled circles with error bars and the model fit is shown by the
histogram. 
{\it Right panel\/}: LXSS model \ion{Fe}{22} $I$(11.92 \AA )/$I$(11.77
\AA ) line ratio as a function of electron density for $T_{\rm e}=12.8$
MK and $T_{\rm bb}=0$, 60, 80, and 100 kK ({\it solid, dotted, short
dashed, and long dashed curves, respectively\/}). Horizontal line and
the dark, medium, and light shaded stripes indicate respectively the
value and the 68\%, 90\%, and 99\% confidence error envelopes of the
line ratio measured in EX Hya.}
\label{fig9}
\end{figure}

\section{Summary}

We have provided a discussion of the density and photoexcitation
sensitivity of the X-ray spectra of collisionally ionized Fe L-shell
ions.  Potentially useful density diagnostics are
identified for
\ion{Fe}{17} and
\ion{Fe}{19}--\ion{Fe}{23}, with the most straightforward being the
\ion{Fe}{17} $I$(17.10 \AA )/$I$(17.05 \AA ) line ratio and the
\ion{Fe}{22} $I$(11.92 \AA )/$I$(11.77 \AA ) line ratio.
Compared to the He-like $R$ density diagnostics, the Fe L-shell density
diagnostics are less sensitive to photoexcitation, hence are particularly
valuable for sources like O stars and cataclysmic variables that are
bright in the ultraviolet. Applying these models to the {\it Chandra\/}
HETG spectrum of the intermediate polar EX~Hya, we have shown that the
observed \ion{Fe}{17} $I$(17.10 \AA )/$I$(17.05 \AA ) line ratio is
consistent with an electron density $n_{\rm e}\gax 2\times 10^{14}~\rm
cm^{-3}$ (or a blackbody photoexcitation temperature $T_{\rm bb}\gax 50$
kK) at the $3\, \sigma $ level, while from the observed \ion{Fe}{22}
$I$(11.92 \AA )/$I$(11.77 \AA ) line ratio, we infer $n_{\rm e}=
1.0^{+2.0}_{-0.5} \times 10^{14}~\rm cm^{-3}$ at the $1\, \sigma $ level
and $n_{\rm e} \gax 2\times 10^{13}~\rm  cm^{-3}$ (or $T_{\rm bb} \gax
100$ kK) at the $3\, \sigma $ level. The high densities are assured in
EX Hya, whose far ultraviolet spectrum has been modeled  as a
$T_{\rm bb} \approx 25$~kK white dwarf \citep{gre97} and as a
$T_{\rm bb} \approx 20$~kK white dwarf with a $T_{\rm bb}\approx 37$~kK
hot spot \citep{mau99}. 
While these results are particular to EX Hya, the Fe L-shell density
diagnostics are applicable to collisionally ionized laboratory (e.g.,
tokamak) or cosmic (e.g., cataclysmic variable) plasmas with electron
temperatures $T_{\rm e}\approx 2$--45 MK and electron densities
$n_{\rm e}\gax 10^{11}~\rm cm^{-3}$.


\begin{theacknowledgments}
Support for this work was provided in part by NASA through {\it
Chandra\/} Award Number DD0-1004B issued by the {\it Chandra\/} X-Ray
Observatory Center, which is operated by SAO for and on behalf of NASA
under contract NAS8-39073. This work was performed under the auspices of
the U.S.~Department of Energy by University of California Lawrence
Livermore National Laboratory under contract No.~W-7405-Eng-48.
\end{theacknowledgments}


\end{document}